\documentclass[a4paper,11pt]{article}
\pdfoutput=1 

\usepackage{jheppub}

\usepackage[T1]{fontenc}

\usepackage{amsmath,amssymb,amsfonts, physics}
\usepackage{mathrsfs}
\usepackage{bbm}
\usepackage{slashed}
\usepackage{graphicx}
\usepackage{verbatim}
\usepackage[T1]{fontenc}
\usepackage[utf8]{inputenc}
\usepackage{url}
\usepackage{hyperref}
\usepackage[english]{babel}
\usepackage{mathbbol}
\usepackage[dvipsnames]{xcolor}
\usepackage[normalem]{ulem}
\usepackage{svg}

\usepackage[backend=bibtex,style=numeric-comp, sorting=none]{biblatex}
\usepackage{xurl}
\bibliography{t-dual.bib}

\numberwithin{equation}{section}
\newenvironment{eqaed}
{\begin{equation}
		\begin{aligned}
		}
		{ 
		\end{aligned}
	\end{equation}
}
\newcommand{\commie}[1]{}

\begin{document}
	
	\author[a,b]{Ivano Basile}
    \author[c,d]{Francesco Ferrarin}
	\author[c,d,e]{Alessia Platania} 
    \author[f]{Matteo Zatti}
	
\affiliation[a]{INFN sezione di Torino, via Pietro Giuria 1, 10125 Torino, Italy}
\affiliation[b]{Jefferson Physical Laboratory, Harvard University, Cambridge, MA 02138, USA}	
\affiliation[c]{Center of Gravity, Niels Bohr Institute, Blegdamsvej 17, DK-2100 Copenhagen \O, Denmark}
\affiliation[d]{Niels Bohr International Academy, Niels Bohr Institute, Blegdamsvej 17, DK-2100 Copenhagen \O, Denmark}
\affiliation[e]{Institute of Physics, University of Graz, Universit\"atsplatz 5, 8010 Graz, Austria}
\affiliation[f]{Max-Planck-Institut für Physik (Werner-Heisenberg-Institut), Boltzmannstrasse 8, 85748 Garching bei München, Germany}

	\emailAdd{ivano.basile@to.infn.it}
    \emailAdd{francesco.ferrarin@nbi.ku.dk}
	\emailAdd{alessia.platania@nbi.ku.dk}
    \emailAdd{zatti@mpp.mpg.de}

\title{Worldsheet functional RG and the $\alpha'$-tower}

\begin{abstract}
{We apply the functional renormalization group to the worldsheet theory of the bosonic string. Imposing the Weyl anomaly cancellation as a condition on the local renormalization group flow, we derive flow equations for the background couplings and the corresponding effective spacetime equations of motion, first in a two-derivative truncation, on a general curved background. We then consider a truncation to all orders in $\alpha'$, which captures only a subset of the possible tensor structures but is analytically tractable, in analogy with $f(R)$-like approximations. Within this truncation, we obtain Tseytlin's spacetime effective action, and we use it to study solutions to all orders in $\alpha'$, including black holes. Our results show that the functional renormalization group can be implemented on the string worldsheet and provide a framework to access the strong-curvature regime. As a longer-term goal, this may allow one to test the scenario of effective asymptotic safety in string theory, in which an approximately scale-invariant regime lies between the low-energy effective field theory and a stringy completion.}
\end{abstract}

\preprint{ MPP-2026-208}

\maketitle

\section{Introduction}\label{sec:intro}

Finding a consistent and empirically adequate quantum theory of gravity~\cite{deBoer:2022zka,Bambi:2023jiz,Basile:2024oms,Buoninfante:2024yth} remains one of the central open problems of theoretical physics. Over the past decades, several approaches have been developed. String theory provides an ultraviolet (UV)-complete framework in which gravity is unified with matter and gauge interactions, and where the problem of perturbative non-renormalizability is cured by the extended nature of the fundamental objects.

In string theory, the gravitational dynamics we observe at low energies emerges as an effective description. Einstein's equations are corrected by an infinite series of higher-curvature terms, suppressed by powers of the string length squared $\alpha'$\footnote{More precisely, $1/\alpha'$ is the tension of the fundamental string, so that $\sqrt{\alpha'}$ can be roughly thought of as the typical length scale.}, and the full tower encodes the extended nature of the string~\cite{Friedan:1980jm,Callan:1985ia,Polchinski:1998rq,Callan:1989nz}. As long as all curvature scales are small compared with $1/\alpha'$, this expansion is under control, and the first few terms suffice to describe low-energy physics at the desired order. However, many of the most interesting physical situations lie precisely outside this regime: the region near cosmological and black-hole singularities, the very early universe, and in general any background where curvature approaches the string scale. In these cases all the terms in the series become comparable, and a truncation at any finite order is not justified. What is needed is a way to resum the $\alpha'$ expansion, or at least a controlled portion of it, so as to follow the physics as the curvature grows. This is challenging. In generic spacetime backgrounds, the perturbative coefficients are known only up to a few orders, and the series itself is expected to be divergent, although asymptotic to the exact answer, rather than convergent. In principle, the curvature problem (and, more generally, the failure of low-energy descriptions at some length/energy scales) is addressed by the worldsheet formulation. In practice, solving the worldsheet theory is only feasible in special classes of backgrounds\footnote{Notably, such backgrounds can be non-geometric, \textit{i.e.} not describable just in terms of a spacetime manifold with background fields. We do not concern ourselves with such settings in this work. Still, it is worth pointing out that the low-energy gravitational theories arising from them share similar features~\cite{Aoufia:2024awo, Basile:2024lcz, Aoufia:2026bau}.}, and in more general cases the analysis effectively amounts to perturbation theory in $\alpha'$. Moreover, some useful statements can be obtained for general geometric backgrounds without expanding in $\alpha'$~\cite{Kutasov:2015eba, Zigdon:2026lkf}. Another issue is that all of these considerations apply at weak coupling, where perturbative string theory makes sense. One can expect that phenomenologically relevant regions of strong curvature also host strong coupling; in this work we will only consider tree-level string theory\footnote{Despite this, there are attempts to gain insight on cosmological-like singularities from the worldsheet~\cite{Arduino:2020axy, Arduino:2022mir, Pesando:2022amk}.}.

The functional renormalization group (FRG) offers a way to address exactly this kind of problem. Rather than computing quantum corrections order by order in a small parameter, the FRG follows the evolution of the effective action as fluctuations are integrated out progressively~\cite{Wetterich:1992yh,Morris:1993qb}. The flow is governed by an exact equation, and any approximation consists in the choice of a subset of operators, rather than in an expansion in the strength of the couplings. As a result, even simple truncations contain contributions from infinitely many orders in perturbation theory. This is the reason for the success of the FRG in contexts where perturbation theory fails, from critical phenomena in statistical physics to the strongly coupled regime of
gauge theories~\cite{Berges:2000ew,Dupuis:2020fhh}.

In gravity, the FRG has become the main tool of the Asymptotic Safety (AS) program~\cite{Percacci:2017fkn,Reuter:2019byg,Bonanno:2020bil,Pawlowski:2020qer,Knorr:2022dsx, Eichhorn:2022gku, Morris:2022btf, Martini:2022sll, Wetterich:2022ncl, Platania:2023srt, Saueressig:2023irs, Pawlowski:2023gym, Bonanno:2024xne,Eichhorn:2026uqj}, which goes back to Weinberg's proposal that a non-trivial UV fixed point could make a quantum field theory of gravity predictive despite its perturbative
non-renormalizability~\cite{Weinberg:1980gg}. Evidence for such a fixed point has since been collected in increasingly sophisticated truncations, including higher-curvature operators~\cite{Reuter:2001ag, Lauscher:2002sq, Codello:2006in, Gies:2015tca, Gies:2016con, Hamada:2017rvn, Knorr:2017fus, Christiansen:2017bsy, Falls:2017lst, Falls:2020qhj, Knorr:2021lll, Knorr:2021slg, Kluth:2022vnq, Baldazzi:2023pep} and matter fields~\cite{Dona:2013qba,Christiansen:2017cxa,Pastor-Gutierrez:2022nki}. At the same time, important questions remain open, such as whether AS can find a fundamental explanation for black hole entropy~\cite{Basile:2025zjc} and, if not, whether AS can still be realized effectively, \emph{e.g.}, as a low-energy approximation to string theory~\cite{deAlwis:2019aud,Basile:2021krk,Basile:2021euh}. It is in this context that confronting the FRG with string theory becomes attractive. String theory provides a setting in which the UV completion is known to exist and where precise consistency conditions hold.

The use of the FRG in a string-theoretic context is not new. In~\cite{Basile:2021krk,Basile:2021euh} it was employed to resum $\alpha'$ corrections at the level of $O(d,d)$-invariant stringy effective actions, and to investigate the interplay between AS and string theory within the framework of effective AS. In the present paper, we take a complementary route. Instead of following the running of the target-space gravitational action, we apply the FRG directly to the worldsheet description of the string, where the fields of the spacetime theory appear as couplings of a two-dimensional quantum field theory and the Regge slope $\alpha'$ plays the role of the loop-counting parameter. In this formulation, the worldsheet RG flow is already the organizing principle that links the string to its low-energy gravitational dynamics, and the FRG is therefore a natural tool to follow this flow beyond the perturbative regime.

Our main goal is to use the FRG to resum $\alpha'$ corrections at the level of the Polyakov worldsheet action~\cite{Polchinski:1998rq}. We derive the flow equation for the effective average action of the worldsheet theory. More precisely, we only consider bosonic strings for simplicity, ignoring the tachyon (and, for the most part, the B-field). Strictly speaking, this is of course inconsistent, but we treat this as a simplified starting point toward non-tachyonic (super)strings. The resulting flow contains some of the known perturbative corrections, but also terms to all orders in $\alpha'$. Requiring that the theory reaches a conformal fixed point provides a resummed version of the spacetime equations of motion; more precisely, they follow by requiring cancellation of the Weyl anomaly, whose contribution at the leading orders in $\alpha'$ yields the critical dimension and the Einstein equations (as well as the equations of motion for other background fields, such as the dilaton), which we subsequently use to derive black hole solutions that are genuinely all-orders in $\alpha'$. Although, as anticipated above, some approximations are involved, our approach is meant as a first qualitative step: we do not claim that the truncation is complete, but we show that the FRG gives a systematic and workable way to go beyond the perturbative series on the worldsheet, allowing us to resum one specific sector of $\alpha'$ corrections. More broadly, this work is part of a program in which ideas and techniques from different approaches to quantum gravity fertilize each other.

The paper is organized as follows. In Sect.~\ref{sec:stage} we provide a brief background overview of the two main ingredients: the worldsheet formulation of perturbative string theory, and the FRG. In Sect.~\ref{sec:ws_frg} we apply the FRG to the bosonic worldsheet in the two-derivative truncation, as a starting point. In Sect.~\ref{sec:ws_frg_beyond2} we extend this treatment to a special kind of all-order truncation, which is amenable to analytic methods. In Sect.~\ref{sec:solutions} we study the spacetime field equations derived by our truncated dynamics, focusing on Schwarzschild-like and Einstein metrics. We provide some concluding thoughts in Sect.~\ref{sec:conclusions}. In App.~\ref{app:B_field} we include the B-field of bosonic string theory within the two-derivative truncation, complementing the analysis of Sect.~\ref{sec:ws_frg}.

\section{Setting the stage}\label{sec:stage}

Before turning to the resummation of $\alpha'$ corrections, we collect the two ingredients on which our analysis rests and which belong to rather different research fields. The first is the worldsheet description of string theory, in which the dynamics of spacetime fields is encoded in a two-dimensional quantum field theory, and the requirement of quantum
consistency of the string (Weyl anomaly cancellation) translates into conditions on the RG flow of its couplings. The second is the FRG, which provides a non-perturbative tool to follow such a flow without relying on an expansion in a small parameter. We review them in turn, with the aim of fixing notation and highlighting the aspects that are relevant for what follows, and refer the reader to the literature for a more complete treatment.

\subsection{Worldsheet formulation of string theory}\label{sec:string_bg}

As anticipated in the Introduction, we consider closed bosonic strings in geometric backgrounds, ignoring the tachyon and the B-field. The former is an important feature of this setting, but we treat the bosonic string as a simplified starting point, with non-tachyonic (super)strings in mind as a future extension.
The string is characterized by an embedding $X: (\Sigma,\gamma) \to (\mathcal{M},G)$ of the worldsheet (parametrized by local coordinates $\sigma^\alpha$) into the spacetime manifold $\mathcal{M}$ (otherwise dubbed target space in this context). Here $\gamma$ and $G$ denote the respective metrics. The starting point is the (Euclidean) Polyakov action,
\begin{eqaed}\label{eq:polyakov_action}
    S_\text{Polyakov} = \frac{1}{4\pi \alpha'} \int d^2\sigma \, \sqrt{\gamma} \left(G_{\mu \nu}(X) \partial_\alpha X^\mu \partial^\alpha X^\nu + \alpha' R^{(2)} \, \phi(X) \right) ,
\end{eqaed}
where $R^{(2)}$ is the Ricci scalar of the worldsheet metric and the dilaton $\phi$ is a scalar field on target space. As we will discuss later, an important feature of Eq.~\eqref{eq:polyakov_action} is that a constant dilaton profile $\phi(X)=\phi_0$ yields a topological term (the Euler characteristic of the worldsheet), and thus the sum over worldsheet topologies in the path integral over the dynamical fields $X$ and $\gamma$ is weighted by~$e^{\phi_0}$, which plays the role of the string coupling. Similarly, when $\phi$ varies slowly enough relative to~$\alpha'$, one can think of $e^\phi$ as a sort of local string coupling.
For our purposes, the key point is that the Polyakov path integral gauges Weyl rescalings $\gamma \mapsto \Omega^2(\sigma) \gamma$. Weyl invariance can be spoiled classically by the dilaton background, and quantum mechanically by the Weyl anomaly~\cite{Callan:1985ia, Callan:1989nz, Tseytlin:1988rr}. Hence, it is paramount that the total variation of the quantum effective action $\Gamma$ under Weyl rescalings vanish. This yields consistency equations for the spacetime background fields, which turn out to be the (classical) equations of motion stemming from a spacetime (Wilsonian) effective action. This logic is not exclusive to strings---in fact, it also holds for point particles~\cite{Dyson:1990:FPM, Bonezzi:2018box, Bonezzi:2024emt}---but in the case of strings the field equations contain higher-derivative corrections weighted by~$\alpha'$. This is due to the extended nature of strings. Moreover, the worldsheet provides a perturbative UV-completion of the dynamics, partially owing to the state-operator correspondence of two-dimensional conformal field theory~\cite{Witten2015WhatEP}.
The Polyakov path integral over $X$ and $\gamma$ can be thought of as two-dimensional quantum gravity on the worldsheet. As such, Weyl and diffeomorphism invariance lead to conformal symmetry, and in particular the corresponding beta functions (including the dilaton contribution) must vanish\footnote{Local Weyl transformations are more general than constant rescalings of the metric. Correspondingly, there is a generalized notion of RG flow where the running of scales is defined locally~\cite{Shore:1986hk}.}.

More precisely, the Weyl anomaly coefficients can be organized into RG beta functions 
\begin{subequations}\label{eq:ws_beta_funcs}
\begin{align}
	&\beta_{\mu \nu}  \equiv k \, \frac{d}{d k} G_{\mu \nu} \, , \qquad \beta^\phi  \equiv k \, \frac{d}{d k} \phi \, ,
\end{align}
\end{subequations}
computed on a flat worldsheet; these differ by derivative terms of the target-space fields, given schematically by~\cite{Tseytlin:1986tt,Tseytlin:1988rr,Tseytlin:2006ak,Bonezzi:2021mub}
\begin{equation}
    \bar \beta^i= \beta^i + \Delta \varphi^i \,,
\end{equation}
where $\varphi^i$ are the target space fields, $\beta^i$ the corresponding beta functions, and $\Delta$ the operator implementing the gauge transformations (possibly with a field-dependent parameter). Therefore, instead of constructing a higher-derivative spacetime theory one could follow the RG flow of the worldsheet theory, and infer the corresponding spacetime dynamics from it. 

A useful consequence of this structure is that the Weyl-anomaly coefficients can be obtained from a covariant functional of the target-space fields. More specifically, the sigma-model RG has been found to possess a gradient-flow structure, such that there exists a local functional of the target-space fields whose variation is related to the Weyl-anomaly coefficients. At a conformal fixed point, this functional is related to the central charge of the worldsheet theory. This structure was identified in the context of the sigma-model c-theorem, and led to the ``central charge action'' developed in~\cite{Tseytlin:2006ak}:
\begin{equation} \label{eq:centralChargeAction}
    \frac{1}{16\pi G_{\rm N}} \int d^Dx \sqrt{-G}e^{-2\phi} \left( {\beta^\mu}_{\mu}-4\beta^\phi \right) .
\end{equation}
Its equations of motion are equivalent to the vanishing of the Weyl-anomaly coefficients order by order in $\alpha'$, up to suitable local field redefinitions and in an appropriate renormalization scheme. These can be recast as
\begin{subequations} \label{eq:beta_conditions}
\begin{align}
	&\beta_{\mu \nu} + 2\alpha' \, \nabla_\mu \nabla_\nu \phi = 0 \, , \\[2mm]
	&\beta^\phi + \alpha' \left(\nabla \phi \right)^2 = 0 \, .
\end{align}
\end{subequations}
Therefore, once the beta functions are known, their spacetime dynamics can be encoded in an effective action in terms of a covariant functional. 

The beta functions are usually computed perturbatively in $\alpha'$, order by order, and it is natural to ask whether the worldsheet RG can be formulated in a way that captures these corrections beyond a finite perturbative order in $\alpha'$. In this work, we investigate this question by means of the FRG, which allows the RG flow to be studied without expanding perturbatively in $\alpha'$. We apply this approach to the two-dimensional nonlinear sigma model parametrized by the target-space metric and the dilaton. As anticipated above, in the main text we set the B-field to zero for simplicity, relegating its FRG analysis (in the two-derivative truncation) to App.~\ref{app:B_field}.

At this point, an important subtlety shows up, when we look for an approach beyond perturbation theory in $\alpha'$: in the quantization of the bosonic string, the vanishing of the beta function of the target space metric, $\beta^G_{\mu\nu}=0$, implies that the dilaton beta function be constant through the Curci-Paffuti relation~\cite{Curci:1986hi}. Specifically, $\beta^{\phi}$ is constrained to be equal to the local central charge, which vanishes in the critical case (once ghosts are included). This relation has been found to persist order by order in the perturbative $\alpha'$ expansion~\cite{Curci:1986hi}, and plays a role akin to a sum rule. This same structure is closely tied to the existence of a spacetime action reproducing the Weyl anomaly cancellation conditions, because it implies that, at a given order in $\alpha'$, the flow is of gradient type~\cite{Tseytlin:2006ak}. In other words, writing the central charge action as a functional whose variation reproduces the anomaly cancellation conditions assumes that this gradient structure persists to all orders. For our purposes, this poses a challenge; approximations of the FRG are not \emph{a priori} guaranteed to preserve the Curci-Paffuti relation, and checking it in any sufficiently complicated truncation might become prohibitive rather quickly. Given this state of affairs, we will not attempt to compute the beta functions on a flat worldsheet and infer $\beta^\phi$ from them. Rather, we shall work directly on a curved worldsheet, despite the technical complications it brings.

\subsection{FRG in a nutshell}\label{sec:frg_bg}

In this section we introduce the FRG and some tools that are typically employed to solve it within certain approximations. See~\cite{Dupuis:2020fhh} for an extensive review of the FRG and its applications.

The FRG implements Wilson's idea of
integrating out fluctuations shell by shell in momentum space at the level of the quantum effective action~\cite{Wilson:1973jj,Polchinski:1983gv,Wetterich:1992yh,Morris:1993qb}. This is done by means of an infrared (IR) regulator, which modifies the bare action in the path integral by a term
$\Delta S_k[\Phi]=\tfrac12\int d^D x \Phi\, R_k\, \Phi$, where $\Phi(x)$ denotes the set of fields in the system, and the kernel $R_k(p^2)$
suppresses modes with $p^2\lesssim k^2$ and vanishes for $p^2\gg k^2$. The resulting scale-dependent effective action $\Gamma_k[\phi]$ is called the Effective Average Action (EAA). It interpolates between the microscopic action, $\Gamma_{k\to\Lambda}\to S$, and the full quantum effective action, $\Gamma_{k\to 0}=\Gamma$. Its dependence on the
RG scale $t=\ln(k/k_0)$ is governed by the exact flow equation~\cite{Wetterich:1992yh,Morris:1993qb}
\begin{equation}\label{eq:wetterich}
	k \partial_k \Gamma_k= \frac{1}{2}\mathrm{STr}\left\{\left(\Gamma_k^{(2)}+\mathcal{R}_k\right)^{-1}k\partial_k \mathcal{R}_k\right\}\,,
\end{equation}
where $\Gamma_k^{(2)}$ is the second functional derivative with respect to the fields $\Phi$, and $\mathrm{STr}$ denotes a (super)trace over all internal indices and momenta. The computation of this trace requires standard heat kernel techniques~\cite{Vassilevich:2003xt}. Eq.~\eqref{eq:wetterich} is exact and has a one-loop structure, but the propagator is the full, field-dependent and scale-dependent one, so that all orders of perturbation theory are resummed. Since Eq.~\eqref{eq:wetterich} is a functional differential equation, it is solved in practice by projecting onto a truncation of $\Gamma_k$, for example a derivative or a vertex expansion. This gives a finite set of beta functions for the couplings. Concretely, one begins by specifying a truncation of the effective average action, corresponding to an expansion in a (possibly infinite) basis of operators,
\begin{equation}
\mathcal{L}=\sum_{n=1}^{N} g_n\mathcal{O}_n\,,
\end{equation}
where the couplings (or, more generally, form factors) are promoted to scale-dependent quantities, $g_n\rightarrow g_n(k)$. The resulting effective average action is supplemented by appropriate gauge-fixing and ghost terms. Substituting this ansatz into Eq.~\eqref{eq:wetterich} and projecting the flow onto the operator basis ${\mathcal{O}_n}$ yields a coupled system of RG equations,
\begin{equation}
k\partial_k g_n=\beta_n(g_1,g_2,\dots)\,,
\end{equation}
where $\beta_n$ denotes the beta function associated with the coupling $g_n$. Since the exact solution of Eq.~\eqref{eq:wetterich} would require including the full, infinite-dimensional theory space, practical calculations necessarily rely on finite-dimensional truncations or on the computation of beta functionals for functions of certain invariants, \textit{e.g.}, for $f(R)$ or $f(R_{\mu\nu\sigma\rho}R^{\mu\nu\sigma\rho})$. The robustness of the resulting fixed-point structure and critical exponents is then assessed by systematically enlarging the truncation and studying the stability, or apparent convergence, of the results~\cite{Denz:2016qks}. 

The FRG was adapted to gravity by Reuter in~\cite{Reuter:1996cp}, who formulated the flow for the effective average action $\Gamma_k[g_{\mu\nu},\bar g_{\mu\nu}]$ using the background-field method, with the regulator built from the covariant Laplacian of the background metric $\bar g_{\mu\nu}$. In the Einstein-Hilbert truncation in Euclidean signature,
\begin{equation}
  \Gamma_k = \frac{1}{16\pi G_k}\int d^Dx\,\sqrt{g}\,\left(2\Lambda_k - R\right)\,,
\end{equation}
the flow of the dimensionless couplings $g_k=k^{D-2}G_k$ and $\lambda_k=k^{-2}\Lambda_k$ possesses a non-Gaussian UV fixed
point in addition to the Gaussian one. A trajectory emanating from such a fixed point is UV complete without being perturbatively renormalizable, which is the essence of Weinberg's AS scenario~\cite{Weinberg:1980gg} and of the concept of non-perturbative renormalizability~\cite{Wilson:1973jj}. The existence of this fixed point has since been found to be stable under a large number of extensions of the truncation, including higher-curvature
terms~\cite{Reuter:2001ag, Lauscher:2002sq, Codello:2006in,Codello:2008vh,Lauscher:2002sq,Benedetti:2009rx,Gies:2015tca, Gies:2016con, Biemans:2016rvp, Hamada:2017rvn, Knorr:2017fus, Christiansen:2017bsy, Falls:2017lst, Falls:2020qhj, Knorr:2021lll, Knorr:2021slg, Kluth:2022vnq, Baldazzi:2023pep} and matter fields~\cite{Dona:2013qba,Biemans:2017zca,Christiansen:2017cxa,Pastor-Gutierrez:2022nki}, and vertex expansions that distinguish the background from the fluctuation field~\cite{Christiansen:2014raa,Denz:2016qks,Pawlowski:2020qer}.

The FRG has a natural application on the worldsheet, which is the
setting of the present paper. In the sigma-model description of the string, the target-space fields (metric, $B$-field, dilaton) appear as coupling functions of a two-dimensional field theory, and the conformal invariance of the worldsheet theory is equivalent to the vanishing of their $\beta$-functionals, which reproduces the target-space equations of motion~\cite{Friedan:1980jm,Callan:1985ia,Callan:1989nz}. At leading order in $\alpha'$ the metric beta function is the Ricci flow, $\beta^G_{\mu\nu}=\alpha' R_{\mu\nu}+\mathcal{O}(\alpha'^2)$, and higher orders in $\alpha'$ correspond to a derivative expansion in target-space curvature.
Because the beta functions obtained from the perturbative expansion in $\alpha'$ are known only to a finite loop order, the question of resumming the series naturally arises. Applying Eq.~\eqref{eq:wetterich} to the worldsheet sigma model~\cite{Codello:2008qq} yields a flow equation whose right-hand side contains contributions from higher orders in $\alpha'$, which reduces to the known perturbative beta functions when expanded in $\alpha'$.

\section{FRG on the worldsheet}\label{sec:ws_frg}    

We now apply the FRG discussed in Sect.~\ref{sec:frg_bg} to the bosonic Polyakov string in $D$-dimensions. As explained there, the construction starts from an ansatz for the worldsheet EAA: as a first step, we consider the simplest possible truncation, namely, the classical Polyakov action where the target space metric and dilaton are promoted to RG-scale-dependent couplings, while the Kalb-Ramond field is set to zero,
\begin{eqaed}\label{eq:truncated_action}
	\Gamma_k = \frac{1}{4\pi \alpha'} \int d^2 \sigma \sqrt{\gamma} \left(G^{(k)}_{\mu \nu}(X) \, \partial_\alpha X^\mu \partial^\alpha X^\nu + \alpha' \, R^{(2)} \, \phi^{(k)}(X) \right) \, .
\end{eqaed}
We shall not specify the details of the ghost sector, since it is not needed in the ensuing discussion; its contribution to the Weyl anomaly can be incorporated later separately. The extension to a non-vanishing $B$-field is discussed in App.~\ref{app:B_field}.

We study the flow by inserting the ansatz~\eqref{eq:truncated_action} into the Wetterich equation~\eqref{eq:wetterich} and projecting the flow onto the theory subspace parametrized by the truncation. From Eq.~\eqref{eq:truncated_action}, the quadratic operator entering the functional trace is
\begin{eqaed}\label{eq:frg_propagator}
	\left(\Gamma_k^{(2)}\right)_{\mu \nu} = \frac{1}{4\pi \alpha'}( G^{(k)}_{\mu \nu}(X) \, (-\Box) - R^{(k)}_{\mu \rho \nu \sigma}(X) \, \partial_\alpha X^\rho \partial^\alpha X^\sigma + \frac{\alpha'}{2} \, R^{(2)} \, \nabla_\mu \nabla_\nu \phi^{(k)}(X)) \, ,
\end{eqaed}
where the d'Alembert operator $\Box$ is defined by the pull-back of the target-space Levi-Civita connection via the embedding $X$. Upon projection onto the truncation, the functional trace can be evaluated via standard heat kernel techniques\footnote{For a detailed account, see~\cite{Vassilevich:2003xt}.}. The resulting projected Wetterich equation reads
\begin{eqaed}\label{eq:frg_beta_functions}
	& \frac{1}{4\pi \alpha'} \int d^2 \sigma \sqrt{\gamma} \left(\beta_{\mu \nu}(X) \, \partial_\alpha X^\mu \partial^\alpha X^\nu + \alpha' \, R^{(2)} \, \beta^{\phi}(X) \right) \\
	& = \frac{1}{4\pi} \int d^2 \sigma \sqrt{\gamma} \left( R^{(k)}_{\mu \nu}(X) \, \partial_\alpha X^\mu \partial^\alpha X^\nu + R^{(2)} \left[ \frac{D-D_c}{6} - \frac{\alpha'}{2} \, \nabla^2 \phi^{(k)} \right] \right) \\
	& + \frac{1}{8\pi} \int d^2 \sigma \sqrt{\gamma} \left( \frac{1}{2} \beta^{\rho \sigma} R^{(k)}_{\mu \rho \nu \sigma }(X) \, \partial_\alpha X^\mu \partial^\alpha X^\nu \right) \, , \\
    & + \frac{1}{8\pi} \int d^2 \sigma \sqrt{\gamma} \left( R^{(2)} \left[ \frac{1}{6} \left( \beta^\mu_\mu - \eta_{\text{gh}}\right) - \frac{\alpha'}{4} \, \beta^{\mu \nu} \nabla_\mu \nabla_\nu \phi^{(k)} \right] \right) \, ,
\end{eqaed}
where $D_c$ denotes the critical dimension and $\eta_{\text{gh}}$ parametrizes the contribution of the $b$ and $c$ ghost anomalous dimensions. One can then extract the beta functions, but they only reproduce the correct one-loop result, given by the second line of Eq.~\eqref{eq:frg_beta_functions}, if $\eta_{\text{gh}} = \beta^\mu_\mu$. Since we have not specified any truncation in the ghost sector, we shall henceforth assume that it leads to the correct one-loop result expanding in $\alpha'$, as required by consistency. 

Combining the resulting RG flow with the universal terms discussed in Sect.~\ref{sec:string_bg}, the Weyl anomaly cancellation conditions can be extracted by adding the universal dilaton terms to the RG beta functions~\cite{Callan:1985ia,Callan:1989nz,Tseytlin:1988rr,Bonezzi:2021mub} and setting the result to zero. We thus obtain
\begin{subequations}\label{eq:frg_eoms}
\begin{align}
	\beta_{\mu \nu} - \alpha' \, R_{\mu \nu} - \frac{\alpha'}{4} \, \beta^{\rho \sigma} R_{\mu \rho \nu \sigma} & = 0 \, , \label{eq:FirstCondition} \\ 
	\beta_{\mu \nu} + 2\alpha' \, \nabla_\mu \nabla_\nu \phi & = 0 \, , \label{eq:SecondCondition} \\
	\frac{D-D_c}{6\alpha'} - \frac{1}{2} \, \nabla^2 \phi - \frac{1}{8} \, \beta^{\mu \nu} \, \nabla_\mu \nabla_\nu \phi + (\nabla \phi)^2 & = 0 \, .
\end{align}
\end{subequations}
Notice that, while the first and the third equations contain the explicit beta functions from the two-derivative truncation, the second equation is simply a consequence of Weyl anomaly cancellation, and is thus exact with respect to the truncation. The first equation determines the beta function of the target space metric implicitly. One can nevertheless solve for $\beta_{\mu \nu}$ in the second equation, yielding the system
\begin{subequations}\label{eq:final_eoms}
\begin{align}
	R_{\mu \nu} + 2 \, \nabla_\mu \nabla_\nu \phi - \frac{\alpha'}{2} \, R_{\mu \rho \nu \sigma} \nabla^\rho \nabla^\sigma \phi & = 0 \, , \\
	\frac{c}{6\alpha'} - \frac{1}{2} \, \nabla^2 \phi + (\nabla \phi)^2 + \frac{\alpha'}{4} \, \left(\nabla \nabla \phi \right)^2 & = 0 \, ,
\end{align}
\end{subequations}
with $c \equiv D - D_c$. This system reproduces the field equations pertaining to the Einstein-dilaton spacetime effective action of string theory in the low-energy limit upon neglecting the last term in each equation. These terms can be intuitively understood as a partial resummation of $\alpha'$ corrections, but, as we have anticipated, one cannot trust Eq.~\eqref{eq:final_eoms} quantitatively \textit{a priori}, since the flow equation employed in this section relies on a truncation of the theory space. One can nevertheless explore the qualitative consequences of Eq.~\eqref{eq:final_eoms}, and to this end black holes and Einstein backgrounds provide simple scenarios where higher-derivative corrections and large-curvature regimes can be investigated.

\subsection{Reconstructing the central-charge effective action}

As anticipated in Sect.~\ref{sec:string_bg}, within the FRG it appears very difficult to recover the dilaton beta function of Eq.~\eqref{eq:frg_eoms} from the consistency of the metric beta function in a flat worldsheet. In principle, this can be done via Curci-Paffuti relations~\cite{Curci:1986hi,Callan:1985ia,Callan:1989nz}, whence~$\beta^\phi$ is constant whenever the RG beta functions on a flat worldsheet vanish\footnote{In fact, in this case $\beta^\phi$ is proportional to the worldsheet central charge.}. However, it could be the case that the two-derivative truncation violates the condition that, whenever $\beta_{\mu \nu} = 0$, $\beta^\phi$ is a constant. This condition is crucial to define a sensible conformal field theory on the worldsheet and a consistent quantization of the Polyakov string via Weyl anomaly cancellation. Keeping these pressing issues in mind, let us nevertheless move on to write down the spacetime effective action following the prescription of~\cite{Tseytlin:2006ak}, which generalizes the two-derivative case~\cite{Callan:1985ia,Callan:1989nz}. Solving Eq.~\eqref{eq:frg_eoms} iteratively, one finds
\begin{subequations}\label{eq:frg_beta_sol}
    \begin{align}
    \beta_{\mu \nu} & = \sum_{n=0}^\infty \frac{{\alpha'}^{n+1}}{4^n} \left( \mathcal{R}^n \cdot R\right)_{\mu \nu} \, , \\
    \beta^\phi & = \frac{D - D_c}{6} - \, \frac{\alpha'}{2} \, \nabla^2 \phi - \, \frac{\alpha'}{8} \, \beta^{\mu \nu} \, \nabla_\mu \nabla_\nu \phi \, ,
\end{align}
\end{subequations}
where $\left(\mathcal{R} \cdot B\right)_{\mu \nu} \equiv R_{\rho \mu \sigma \nu} B^{\rho \sigma}$ defines the action of the operator $\mathcal{R}$ on symmetric rank-two tensors, whose operator power is denoted by $\mathcal{R}^n$. The string-frame effective action of~\cite{Tseytlin:2006ak} then reads
\begin{eqaed}\label{eq:tseytlin_action}
    S = \frac{1}{16\pi G_{\rm N}} \int d^Dx \sqrt{-G} \, e^{-2\phi} \left( {\beta^\mu}_\mu - \, 4 \, \beta^\phi \right) ,
\end{eqaed}
and contains infinitely many higher-derivative terms. In principle, if we had the exact beta functions, varying Eq.~\eqref{eq:tseytlin_action} would yield the $\alpha'$-exact counterpart of Eq.~\eqref{eq:frg_eoms}. In order to simplify the ensuing analysis, we deform the action~\eqref{eq:tseytlin_action} so that the only dilaton dependence in the effective action be given by the second term in $\beta^\phi$ in Eq.~\eqref{eq:frg_beta_sol}. We then consider
\begin{eqaed}\label{eq:tseytlin_actionV2}
    S = \frac{1}{16\pi G_{\rm N} }\int d^Dx \sqrt{-G} \, e^{-2\phi} \left(- \, \frac{2c}{3\alpha'} + 4 \left(\partial \phi\right)^2 + \frac{1}{\alpha'} \, {\beta^\mu}_\mu - \, \frac{1}{4\alpha'} \, \beta^{\mu \nu} \, \beta_{\mu \nu} \right) \, .
\end{eqaed}
The difference between \eqref{eq:tseytlin_action} and \eqref{eq:tseytlin_actionV2} is a term linear in the combination \eqref{eq:SecondCondition}. The approximate actions are then equivalent provided that we enforce such constraint, for instance with a Lagrange multiplier. In principle, if we had the exact beta functions, varying Eq.~\eqref{eq:tseytlin_action} would yield the $\alpha'$-exact counterpart of Eq.~\eqref{eq:frg_eoms}. One can then induce a deformation linear in the Weyl anomaly cancellation conditions performing a simple field redefinition $\varphi' = \varphi + \delta \varphi$ of the fields which preserves the two-derivative structure. However, such a deformation would not come alone, but as part of a tower of operators associated with higher powers $\delta\varphi$. One can then interpret \eqref{eq:tseytlin_actionV2} as the result of embedding the operators appearing in the approximated action \eqref{eq:tseytlin_action} into the exact theory, keeping track of the field redefinition of such operators discarding non-linear terms and finally projecting out all operators not originally included in the approximated action \eqref{eq:tseytlin_action}. We will comment further on whether this procedure introduces some pathologies when discussing some classes of solutions in Sect.~\ref{sec:solutions}.

It is important to stress again that this reconstruction should not be identified with a derivation of exact string effective action, since the beta functions entering the prescription of~\cite{Tseytlin:2006ak} are obtained via a finite FRG truncation of the full theory space and therefore inherit its limitations. The natural next step going forward is to enlarge the truncation beyond the two-derivative order, thereby allowing a larger portion of theory space to contribute to the RG flow and, as a direct consequence, to the spacetime effective action.

There are two natural ways to pursue such an extension. The first possibility is to retain the full set of couplings and include all additional operators up to a given higher-derivative truncation order. This is the more complete strategy, and particularly well-suited if the ultimate goal is that of reconstructing the target-space effective action. In this setup, one can study solutions to the equations of motion, but also, upon adopting the prescription of~\cite{Tseytlin:2006ak}, compare the resulting Wilson coefficients with information stemming from string scattering amplitudes (assuming the validity of the central-charge action for all-order $\alpha'$). The price to pay is a rapidly increasing computational complexity, as the number of independent operators and couplings grows considerably with the truncation order. 

An alternative is to restrict the calculation to a smaller and simpler sector of theory space. In particular, by constraining the dilaton to be constant, the corresponding term becomes topological:
\begin{equation}
\frac{1}{4\pi}\int d^2\sigma\sqrt{\gamma}\phi_0 R^{(2)}
=\phi_0\chi(\Sigma)\,,
\end{equation}
and therefore carries no local information about the dynamics. As a consequence, one can restrict to a flat worldsheet without loss of generality. This removes a significant number of structures from the calculation, simplifying it significantly and allowing to extend the truncation to substantially higher-derivative orders. 

The simplification comes at the cost of restricting the portion of theory space considered, since operators involving derivatives of the dilaton are not accessible, and therefore the calculation is to be understood as a controlled projection of the full theory. As such, it does not provide a complete off-shell effective action in a specified operator basis, and the resulting coefficients should therefore not be interpreted as the complete set of Wilson coefficients. As such, they cannot necessarily be compared with those from string scattering amplitudes, which are invariant under local field redefinitions. On the other hand, the advantage is that we can access a larger number of higher-derivative operators and study therein solutions to the equations of motion with a high degree of isometry.

In what follows we pursue this second strategy, but implement the extension in a functional derivative-expansion form, allowing the dependence of the effective average action on the tensorial structures to be encapsulated by a general function. This way, we can include an infinite subset of higher-derivative terms, all the while dealing with a tractable and computationally feasible flow.

\section{FRG beyond the two-derivative truncation}\label{sec:ws_frg_beyond2}

We now consider a generalization of the truncation employed in Sect.~\ref{sec:ws_frg}. Rather than fixing the form of the effective average action to match the classical Polyakov action, we allow the dependence on the two building blocks of Eq.~\eqref{eq:truncated_action} to be an arbitrary analytic function. In practice, inspired by infinite-dimensional truncations such as $f(R)$ in the context of gravity, we make the ansatz
\begin{equation}\label{eq:gen_truncation}
    \Gamma_k = \frac{1}{4\pi} \int d^2\sigma \sqrt{\gamma} F_k\left(G_{\mu\nu}^{(k)}\partial X^\mu \cdot \partial X^\nu/ \alpha',\, R^{(2)}\phi^{(k)}\right) ,
\end{equation}
with the Kalb–Ramond field set to zero throughout. The two-derivative truncation of Eq.~\eqref{eq:truncated_action} is recovered for $F_k(x,y)=x+y$. The truncation~\eqref{eq:gen_truncation} is of course not the most general one, but it has the advantage of being analytically tractable.

To determine the flow for $F_k$, we can exploit the structure of the sigma-model beta functions. In particular, the beta function $\beta^{(G)}_{\mu \nu}$ of the target-space metric is dilaton-independent, whereas the dilaton beta function $\beta_{\phi}$ is affine in $\phi$: $\beta_\phi=A[G]\phi+B[G]$, where $A[G]$ a differential operator and $B[G]$ a scalar functional constructed from target-space curvature invariants~\cite{Bonezzi:2021mub, Tseytlin:1988rr}. This structure implies that the two arguments of $F_k$ can be isolated by an appropriate choice of background.

Concretely, we can identify a family of backgrounds parametrized by a constant 2-d vector $v$, related to the pullback derivative $\partial_{\alpha}X^{\mu}$, and the worldsheet curvature $R^{(2)}$, which determine the embedding $X^\mu(\sigma)$ and the worldsheet metric $\gamma_{\alpha \beta}$. Two relevant limits are then: 
\begin{enumerate}
    \item \textit{Target-space metric sector}: set $R^{(2)}=0$ while retaining a non-vanishing $v$. The dilaton coupling drops out, and the target-space metric contribution is isolated.
    \item \textit{Dilaton sector}: set $v=0$, such that the embedding is constant in the worldsheet, and retain a non-vanishing worldsheet curvature $R^{(2)}$.
\end{enumerate} 
This division amounts to moving within each family of backgrounds while keeping the truncation ansatz fixed. This allows us to isolate one contribution or the other, which can then be combined to reconstruct the full RG flow of $F_k$. 

\subsection{Target-space metric sector}
\label{sec:metric_sector}
Restricting to a flat worldsheet metric, $F_k$ reduces to a function of a single argument:
\begin{equation}
    \mathcal{L}= \frac{1}{\alpha'}G_{\mu\nu}^{(k)}(X)\partial X^\mu \cdot \partial X^\nu \,.
\end{equation}
Before proceeding, we note the truncation conditions on $F_k$: since the two-derivative truncation must be recovered at leading order, $F'_k(0)=1$; moreover, $F_k(0)=0$, which implies $\dot F_k(0)=0$. These will be useful to fix the solution below.

We again employ the background field method with exponential splitting to preserve covariance, working in Riemann normal coordinates centered on the background $X_0.$ Expanding to quadratic order in the fluctuation $\xi$, we have 
\begin{equation}\label{eq:quad_F}
    F_k(\mathcal{L})|_2=\frac{1}{2} F_k''(\mathcal{L}_0) \mathcal{L}_1^2+ F_k'(\mathcal{L}_0) \mathcal{L}_2 \,,
\end{equation}
with 
\begin{equation}
\begin{split}
    \mathcal{L}_0=& \frac{1}{\alpha'}G_{\mu\nu}(X_0)\partial  X_0^\mu \cdot \partial X_0^\nu\,, \\
    \mathcal{L}_1=& \frac{2}{\alpha'} G_{\mu\nu}(X_0)\partial  X_0^\mu \cdot \nabla  \xi^\nu\,, \\
    \mathcal{L}_2=& \frac{1}{\alpha'}(G_{\mu\nu}(X_0)\nabla  \xi^\mu \cdot \nabla  \xi^\nu-R_{\mu \rho \nu \sigma}\partial  X_0^\rho \cdot \partial  X_0^\sigma \xi^\mu \xi^\nu)\,. 
\end{split} 
\end{equation}
It is useful to introduce the target-space vielbeins $e^\mu_m$ trivializing the metric, and write $\xi^\mu=e^\mu_m \chi^m$, so that $\mathcal{L}_1=2/\alpha' e_{\mu m}\partial X^\mu \cdot D\chi^m$.

We now choose the background embedding such that the pullback derivative is constant on the worldsheet, imposing $\partial_\alpha X_0^\mu=V^\mu_\alpha$ for some constant $V^\mu_\alpha$. This is an overdetermined system of PDEs for $X_0^\mu$. Integrability requires the Frobenius condition that mixed partial derivatives commute, which implies $\partial_{[\alpha}V_{\beta]}^\mu=0$. We satisfy this by taking $V^\mu_\alpha$ to be rank-one,
\begin{equation}\label{eq:const_deriv}
    \partial_\alpha X_0^\mu = e^{\mu}_m u_{\alpha} v^m \,,
\end{equation}
where $u_\alpha$ is a constant covector on the worldsheet and $v^m$ is a constant vector in the flat target-space frame. Since the worldsheet is topologically $\mathbb{R}^2$, this ensures that a global solution $X_0(\sigma)$ exists. 

The choice~\eqref{eq:const_deriv} has two immediate consequences. First, the coefficient of $D\chi$ in $\mathcal{L}_1$ is constant, so only constant terms appear in the quadratic action. Second, and most importantly, $\mathcal{L}_0=u^2 v^2/\alpha'$, so the argument of $F', \, F''$ appearing in Eq.~\eqref{eq:quad_F} is constant throughout. Passing to flat frame indices, the quadratic action reads
\begin{equation}
     \frac{1}{4 \pi \alpha'} \left (F'_k D\chi^m \cdot D\chi^n \delta_{m n}+2F_k''u^\alpha u^\beta \frac{v_m v_n}{\alpha'}  D_\alpha \chi^m D_\beta \chi^n - F_k' R_{m a n b} u^2 v^a v^b \chi^m \chi^n \right )
\end{equation}
where the primes are understood to be evaluated at $u^2 v^2/\alpha'$. Since the curvature is zero and there are no topological obstructions, the spin connection is pure gauge and can be set to zero. The Hessian of the effective average action then reads
\begin{equation}
    \left ( \Gamma^{(2)}_k \right)_{mn} = \frac{1}{4\pi \alpha'} \left ( F'_k \delta_{mn} (-\Box) - 2 F''_k (u \cdot \partial)^2 \frac{v_m v_n}{\alpha'} - F'_k  R_{m a n b} u^2 v^a v^b \right ) \,.
\end{equation}
We now turn to the evaluation of the Wetterich equation in this sector. We adopt the Litim regulator
\begin{equation}\label{eq:cutoff}
    \mathcal{R}_k = \frac{1}{4 \pi \alpha'} G_\mu{}^\rho R_k(-\Box)_{\rho\nu}\,, \qquad R_k(z) = (k^2 - z)\,\Theta(k^2 - z)\,,
\end{equation}
which renders all momentum integrals algebraic and allows for explicit analytical control. Note that since the target-space metric carries a $k-$dependence, its $t-$derivative contributes a term proportional to $\beta^{(G)}$ on the right-hand side of the Wetterich equation. Since the worldsheet is flat and the spin connection is pure gauge, the only nonvanishing heat kernel coefficient is $b_0=\mathbb{1}$.

Since the right-hand side of the Wetterich equation contains a ratio, we can go to flat frame indices, and we pass to Fourier space, owing to the fact that the spin connection is pure gauge. We further use the freedom on $u_\alpha$ to fix $u^2=1$. As such, it defines a preferred direction on the worldsheet, and it is convenient to write the momentum integral in polar coordinates. The Wetterich equation reads
\begin{equation}
\begin{split}
    \dot F_k + F'_k \beta_{v v } &= \frac{\alpha'}{4\pi} \int dz \,d\theta \text{tr}\left ( \frac{\beta^{(G)} R_k+\dot R_{k}}{z F'_k \mathbb{1} + 2 F''_k z \cos^2\theta \, \frac{v v}{\alpha'} + \mathbb{E}+R_{k}} \right) \,,
\end{split}
\end{equation}
where $\beta_{vv}=\beta_{ab}v^a v^b$ and $\mathbb{E}_{ab}=- F'_k  R_{m a n b}v^m v^n$, and $R_k=R_k(z)$ is understood. The trace is evaluated by expanding the matrix-valued denominator in powers of the rank-one tensor $v v^ T$. Using $(v v^ T)^k=(v^2)^{k-1} vv^ T$ for the trace one finds, \textit{e.g.},
\begin{equation}
    \begin{split}
    \text{tr}\left( \beta f(vv) \right) &= \text{tr}\Big( \beta \sum_{k \geq0} a_k (v v^ T)^k \Big) = a_0 \text{tr}\beta + \sum_{k \geq1} a_k (v^2)^{k-1} \text{tr}\left( \beta v v \right) \\
    & = a_0 \text{tr}\beta + \frac{f(v^2)-f(0)}{v^2} \beta_{vv} \,, \\[2mm] \text{tr}\left( \mathbb{E} f(vv) \right) &=\sum_{k \geq1} a_k \text{tr}\left( \mathbb{E} (vv)^k \right)=0 \,,
\end{split}
\end{equation}
where the second line follows from antisymmetry of the Riemann tensor. Applying these identities and inserting the Litim regulator, the Wetterich equation becomes
\begin{align}
    \dot F_k + F'_k \beta_{v v } &= \frac{\alpha'}{2} \int_0^{k^2} dz \left [ \int \frac{d\theta}{2 \pi}\frac{2k^2}{(F'_k-1)z+k^2+2F''_kz \frac{v^2}{\alpha'} \cos^2\theta}- \frac{2 k^2 }{((F'_k-1)z+k^2)^2}\text{tr}\mathbb{E} \right. \nonumber \\[1mm] & \hspace{-0.5cm} \left. +\frac{k^2 -z}{v^2} \beta_{vv} \left ( \int \frac{d\theta}{2 \pi}\frac{1}{(F'_k-1)z+k^2+2F''_kz \frac{v^2}{\alpha'} \cos^2\theta} - \frac{1}{(F'_k-1)z+k^2}\right ) \right.  \\[1mm]  &  \hspace{-0.5cm} \left. -\frac{k^2 -z}{((F'_k-1)z+k^2)^2} \text{tr}\beta \mathbb{E} \right ] \nonumber \,,
\end{align}
where the term proportional to $\text{tr}\beta$ has been dropped due to the truncation condition, since tr$\beta$ is a function of curvature invariants only but $\dot F_k$ is not. Matching left- and right-hand sides, we read off separate conditions for $F'_k$ and $\dot F_k$. The one-loop result requires the coefficients of $\beta_{vv}$ and $R_{vv}$ to be the same. Imposing this condition and introducing $\varphi=F_k'-1$ with the rescaling $y=z/k^2$, one finds
\begin{equation}\label{eq:cond_F'}
    \begin{split}
        & \varphi +1 - \int_0^1  \frac{dy}{(\varphi y+1)^2}  \\ & \hspace{0.5cm} = \frac{\alpha' k^2}{2 v^2} \int_0^1 dy (1-y) \left ( \int \frac{d\theta}{2 \pi}\frac{1}{\varphi y+1+2\varphi' y \frac{v^2}{\alpha'} \cos^2\theta} - \frac{1}{\varphi y+1}\right )  \,,
    \end{split}
\end{equation}
from which it follows that
\begin{equation}\label{eq:eom with varphi}
    \beta_{vv}=\alpha' R_{vv} + \frac{\alpha'}{2} R_{a v b v} \beta^{ab} \frac{\int_0^1dy \frac{1-y}{(\varphi y+1)^2}}{\int_0^1dy \frac{1}{(\varphi y+1)^2}} 
\end{equation}
The ratio of integrals appearing in Eq.~\eqref{eq:eom with varphi} must be independent of the background, since it multiplies $\text{Riem} \cdot \beta$, and must be a pure number. $\varphi$ depends on $v^2$ and $k$, and since the integrals evaluate straightforwardly, yielding 
\begin{equation}
    \frac{1}{2} \frac{1+\varphi}{\varphi^2} (\varphi-\log(1+\varphi))=\text{constant} \,.
\end{equation}
This implies that $\varphi$ is constant. The condition $F_k'=1+\dots$, then immediately gives $\varphi=0$, which is consistent with Eq.~\eqref{eq:cond_F'}. At $\varphi=0$, the ratio of integrals evaluates to $1/2$ directly, so that, substituting back,
\begin{equation}
    \beta_{vv}=\alpha'R_{vv} + \frac{\alpha'}{4} R_{abcd}\beta^{ac}v^b v^d \,.
\end{equation}
Starting from the requirement on the truncation yields back the same result from the two-derivative truncation. 

\subsection{Dilaton sector}
\label{sec:dilaton_sector}
We now study the complementary limit of the background family, setting $v=0$ and retaining a constant non-vanishing worldsheet curvature $R^{(2)}$. For notational simplicity, we will refer to it as $r=R^{(2)}$. With $v=0$, the embedding is constant on the worldsheet, $X^{\mu}(\sigma)=X^{\mu}_0$, and the complicated structure involving the target-space metric is highly simplified. In this limit, $F_k$ is a function of both arguments as in Eq.~\eqref{eq:gen_truncation}, but the structure is considerably simpler. The quadratic action reads
\begin{equation}
     \frac{1}{\alpha'}F_k^{(1,0)}G_{\mu\nu}\nabla  \xi^\mu \cdot \nabla  \xi^\nu + \frac{r}{2} F_k^{(0,1)} \xi^{\mu} \nabla_{\mu}\nabla_{\nu}\phi \xi^{\nu} + \frac{r^2}{2} F_k^{(0,2)} \xi^{\mu} \nabla_{\mu}\phi \xi^{\nu} \nabla_{\nu}\phi \,, 
\end{equation}
where the partial derivatives of $F_k$ are understood to be evaluated at $(0, r \phi_0)$. Since the Polyakov action depends on the dilaton only through its derivatives and its constant mode $\phi_0^{(k)}$ is encoded in $g_s$, we can shift $\phi$ and set $\phi_0^{(k)}=0$ without loss of generality, consistently for all $k$ since $X_0$ is constant by construction and it does not depend on $\sigma$. 

Restricting to a constant embedding in the worldsheet simplifies the pictures significantly. The Hessian of the effective average action reduces to a minimal kinetic operator plus an endomorphism,
\begin{equation}
    \Gamma_k^{(2)}=\frac{1}{4\pi \alpha'} \left (-F_k^{(1,0)} \Box + r \alpha' \mathbb{E}\right)\,,
\end{equation}
where the factor $r$ has been made explicit so that it acts as a bookkeeping parameter for the heat kernel expansion. The left hand side of the Wetterich equation reads
\begin{equation}
    \frac{1}{4 \pi \alpha'} \left ( \dot F_k + F_k^{(0,1)} \alpha' r \beta_\phi \right ) \,,
\end{equation}
and we therefore retain only terms up to order $r$ in the right hand side. We adopt the same cutoff regulator as in the metric sector~\eqref{eq:gen_truncation}. Expanding the supertrace via the heat kernel, the right hand side of the Wetterich equation reads
\begin{equation}
\begin{split}
    \frac{1}{8 \pi} \Bigg  ( & Q_1(f_1) D + Q_1(g_1) \text{tr}\beta^{(G)} - \alpha'r Q_1(f_2) \text{tr}\mathbb{E} \\ & - \alpha' r Q_1(g_2) \text{tr}[\mathbb{E} \beta^{(G)}]  + \frac{\alpha'r}{6}D Q_0(f_1)  + \alpha' r Q_0(g_1) \text{tr}\beta^{(G)} \Bigg) \,,
\end{split}
\end{equation}
where we introduced 
\begin{equation}
    f_n=\frac{\dot R_k}{(-F_k^{(1,0)} \Box + R_k)^n }, \qquad g_n=\frac{R_k}{(-F_k^{(1,0)} \Box + R_k)^n }\,.
\end{equation}
As in Sec.~\ref{sec:ws_frg}, the ghost sector contributes as usual with $\eta_{\text{gh}}$, and by shifting $D \to D - D_c$ in the constant term. Matching the coefficient of $r$ on both sides yields a condition on $F^{(0,1)}_k$,
\begin{equation}
    \frac{2}{\alpha'}F_k^{(0,1)} \beta_\phi =  \left (- Q_1(f_2) \text{tr}\mathbb{E} - Q_1(g_2) \text{tr}[\mathbb{E} \beta^{(G)}] + \frac{D}{6} Q_0(f_1) + Q_0(g_1) \text{tr}\beta^{(G)} \right) \,.
\end{equation}
$\beta_{\phi}$ is a local functional of target-space curvature invariants alone, and it does not depend on $k$, $R^{(2)}$, or $\phi$ explicitly. Consistency of the two sides therefore requires that the $F_k^{(0,1)}$ dependence of the r.h.s. drop out, which is only possible if $F_k^{(0,1)}$ is constant. The two-derivative truncation finally sets $F_k^{(0,1)}$ to be exactly unity, and thus the two-derivative truncation is recovered. 

\subsection{Generic background and $\alpha'$}
The two backgrounds considered above isolate two limits of the family of $(v,R^{(2)})$, and both limits land on the two-derivative truncation. To conclude that $F_k(x,y)=x+y$ identically within our approximation, rather than only along these two limits, we now consider a generic background with both $v\neq0$ and $R^{(2)}\neq0$ simultaneously. We start focusing on the target-space metric and its beta function. In addition to the structures already considered in Sect.~\ref{sec:metric_sector}, the Hessian of $\Gamma_k$ now contains terms carrying an overall factor of $R^{(2)}$; from the form of the left-hand side of the Wetterich equation, such pieces do not contribute to the beta function of $G^{(k)}$, since they do not belong to the metric-projected right-hand side. The pieces that do feed into $\beta^{(G)}$ are therefore the same tensor structures identified in Sect.~\ref{sec:metric_sector}, the only difference being that $F_k'$ and $F_k''$ are not constant. The evaluation of the right-hand side of the Wetterich equation then goes as before, up to additional pieces involving worldsheet derivatives of $F_k'$, $F_k''$; these, too, come with explicit overall factors of $R^{(2)}$, so by the same power-counting argument they cannot contribute to $\beta^{(G)}$. The projected supertrace is hence identical in form to the one computed in Sect.~\ref{sec:metric_sector}, so the condition $\partial_x F_k = 1$ obtained there holds in fact for all $(x,y)$, and not just for $y=0$. Integrating in $x$,
\begin{equation}
    F_k(x,y) = x + f_k(y)\,,
\end{equation}
for some function $f_k$. 

We now turn to the dilaton beta function and set $x=0$, \textit{i.e.} $ v^2 =0$, so that the dilaton field is constant, and so are the derivatives of $F_k$. We can therefore repeat the dilaton-sector analysis of Sect.~\ref{sec:dilaton_sector} as before, and apply it to $F_k(0,y) = f_k(y)$. Again, $F_k^{(0,1)}=\partial_y f_k$ is forced to be constant, and the truncation condition $f_k(0)=0$ fixes it to unity. This confirms, even for a generic background, the two-derivative truncation within this ansatz,
\begin{equation}
    F_k(x,y) = x + y\,.
\end{equation}

Finally, we discuss the running of $\alpha'$, which up to now we had treated as scale-independent; the ansatz for the EAA given in Eq.~\eqref{eq:gen_truncation} is well suited to account for this running. We do this by absorbing it into a redefinition of the target-space metric itself, $G_{k,\,\mu \nu}=G_{\mu \nu}^{(k)}/\alpha_k'$, and taking $G_{k}$ to be the fundamental $k$-dependent quantity entering the argument of $F_k$ in Eq.~\eqref{eq:gen_truncation}. In other words, we are expressing the flow in string units. With this identification, the analysis of Sects.~\eqref{sec:dilaton_sector},~\eqref{sec:metric_sector} goes through unchanged, since none of it relies on $\alpha'$ being $k$-independent. As a consequence, accounting for the running of $\alpha'$ does not change these conclusions: the two-derivative truncation is still recovered, and the running of $\alpha'$ can be absorbed into the normalization of $G_k$.

\section{Spacetime solutions to the EOMs}\label{sec:solutions}

In this section we derive some backgrounds solving the equations of motion (EOMs) of the action~\eqref{eq:tseytlin_actionV2} we constructed and comment on their features. 

\subsection{Perturbative regime}

Let us start by showing that the action \eqref{eq:tseytlin_actionV2} contains an infinite tower of higher-derivative corrections encoded in the $\beta_{\mu\nu}$ dependency of the last two terms. $\beta_{\mu\nu}$ is indeed defined implicitly by Eq.~\eqref{eq:FirstCondition}, which can be conveniently written as 
\begin{equation} \label{eq:betaG}
    Q_{\alpha \beta}\,{}^{\mu\nu} \beta_{\mu\nu} = \alpha'R_{\alpha \beta} \,, \qquad Q_{\alpha \beta}\,{}^{\mu\nu} = \delta_{\alpha}^{(\mu}\delta_{\beta}^{\nu)} - \frac{\alpha'}{4} R_{\alpha}{}^{\mu}{}_{\beta}{}^{\nu}\,,
\end{equation}
where we introduce an operator $Q$ acting on rank 2 symmetric tensors. The operator $Q$ can be inverted point-wise, provided that the Riemann tensor spectrum does not admit $4/\alpha'$ as eigenvalue. The inverse of $Q$ is called its resolvent. It does not have a simple expression, however it admits an analytic expansion provided that the spectral radius of the Riemann tensors (\textit{i.e.} the norm of the largest eigenvalue) is smaller than $4/\alpha'$. In this case one can write 
\begin{equation}
    (Q^{-1})_{\alpha \beta}\,{}^{\mu\nu} = \sum_{k\ge 0} \left(\frac{\alpha '}{4}\right)^k (R^k)_{\alpha \beta}\,{}^{\mu\nu} \,,
\end{equation}
where 
\begin{equation}
    (R^0)_{\alpha \beta}\,{}^{\mu\nu} = \delta_{\alpha}^{(\mu}\delta_{\beta}^{\nu)} \,, \qquad (R^{k+1})_{\alpha \beta}\,{}^{\mu\nu} = R_{\alpha}{}^{\rho}{}_{\beta}{}^{\sigma}(R^{k})_{\rho \sigma}\,{}^{\mu\nu} \,.
\end{equation}
We can then use the series expansion of the resolvent to write the last two terms of the action \eqref{eq:tseytlin_actionV2} as a series of powers of the Riemann curvature
\begin{equation}
     \frac{1}{\alpha'} \, {\beta^\mu}_\mu - \, \frac{1}{4\alpha'} \, \beta^{\mu \nu} \, \beta_{\mu \nu} = R - \frac{\alpha '}{4}
    R_{\mu\nu} \, R^{\alpha\beta}\,\sum_{k\ge 1} k \left(\frac{\alpha '}{4}\right)^{k} (R^k)_{\alpha \beta}\,{}^{\mu\nu} \,.
\end{equation}

We want now to determine for a simple class of solutions which is the characteristic scale at which we leave the perturbative regime. Let us consider the ansatz for a static and spherically symmetric configuration
\begin{equation}
    ds^2 = - N^2 f dt^2 + f^{-1} dr^2 + r^2 d\Omega_{d-2} \,.
\end{equation}
The operator $Q$ generically maps symmetric rank 2 tensors to symmetric rank 2 tensors. It is not difficult to verify that for this background, $Q$ maps the subspace of diagonal tensors to itself. Given that the Ricci tensor is diagonal, we can therefore represent $Q$ as a $D\times D$ operator acting on vectors. Explicitly we have
\begin{equation}
Q_{\mu \nu}^{(d)}\bigg \rvert_{\rm diag} = \delta_{\mu \nu} - \frac{\alpha'}{2} (R_\mu {}^\nu{}_\mu{}^\nu)\,.
\end{equation}
For the case of interest, namely the computation of $\beta_{\mu\nu}$, it is not difficult to verify that the angular components are not independent and Eq.~\eqref{eq:betaG} reduces to
\begin{equation}
    (\beta_{tt}, \beta_{rr}, \beta_{\theta\theta})^T = \alpha' Q^{(3)}{}^{-1} (R_{tt},R_{rr}, R_{\theta\theta})^T \,,
\end{equation}
where the $Q^{(3)}$ operator acting on the Ricci can be described via a $3\times3$ matrix whose explicit expression is
\begin{equation} \label{eq:Q3d}
    Q^{(3)} = \mathbf{1}_3 -   \frac{\alpha'}{2} \begin{pmatrix}
0 & R_0{}^1{}_0{}^1 & \omega_{mn}  R_0{}^m{}_0{}^n  \\
R_1{}^0{}_1{}^0 & 0 & \omega_{mn} R_1{}^m{}_1{}^n  \\
R_2{}^0{}_2{}^0 & R_2{}^1{}_2{}^1 &  \omega_{mn} R_2{}^m{}_2{}^n 
\end{pmatrix} \,,
\end{equation} 
with $\omega_{mn}$ the metric of the $(D-2)$-dimensional sphere of unit radius. The radius $r^*$ at which the perturbative expansion breaks is then fixed as the maximum radius for which $\operatorname{ker} Q^{(3)} \ne 0$.

In order to have some explicit result we specialize to the case of Schwarzschild black hole. We pick
\begin{equation}
    N(r) = 1\,, \qquad f(r ) = 1 - \frac{r_S^{D-3}}{r^{D-3}} \,.
\end{equation}
The eigenvalues of \eqref{eq:Q3d} are then
\begin{equation}
    \left(1 , 1 - \frac{D(D-3)}{4}\frac{\alpha'r_S^{d-3}}{r^{D-1}} , 1 + \frac{(D-2)(D-3)}{4}\frac{\alpha'r_S^{D-3}}{r^{D-1}}\right) \,,
\end{equation}
where we introduced the Schwarzschild radius $r_S$. The perturbative expansion of $Q^{-1}$ in powers of $\alpha'$  is broken for $r < r^*$ with
\begin{equation}
    r^* = \left(\frac{D(D-3)}{4} \alpha' r_S^{D-3} \right)^{1/(D-1)} \,.
\end{equation}
Observe that for black holes larger than the string scale, the scale $r^*$  at which the perturbative expansion breaks down lies inside the event horizon, whereas for black holes smaller than the string scale, it lies outside the event horizon.

\subsection{Einstein solutions of the variational EOMs}

We are now interested in understanding which backgrounds are  solutions of the theory~\eqref{eq:tseytlin_actionV2}. The main difficulty we encounter solving the EOMs is evaluating $\beta_{\mu\nu}$. Our strategy is then to select a class of backgrounds for which $\beta_{\mu\nu}$ can be evaluated easily and then determine the further constraints imposed by EOMs.\footnote{More complicate solutions can be constructed applying recursively the machinery of \cite{Cano:2018qev, Cano:2019ycn,Cano:2021nzo,Ortin:2021win,Cano:2022tmn,Zatti:2023oiq,Massai:2023cis,Zatti:2024vbz}. We leave this analysis for future work.} We consider an Einstein spacetime
\begin{equation} \label{eq:Einstein}
    R_{\mu\nu} = h \, g_{\mu\nu} \,.
\end{equation}
where the Bianchi identity for the Ricci forces $h$ to be a constant in $D > 2$. It is not difficult to verify that hold the identities
\begin{subequations} \label{eq:identitiesRicci}
\begin{align}
& Q_{\mu\nu}{}^{\alpha \beta} g_{\alpha \beta} = \left(1-\frac{\alpha'}{4}h \right) g_{\mu \nu} \,, \qquad (Q^{-1})_{\mu\nu}{}^{\alpha \beta} g_{\alpha \beta} = \left(1-\frac{\alpha'}{4}h \right)^{-1} g_{\mu \nu} \,, \\
& \beta_{\mu \nu} = \alpha' (Q^{-1})_{\mu\nu}{}^{\alpha \beta}  R_{\alpha \beta} = x g_{\mu \nu} \,, \qquad x = \alpha' h \left(1-\frac{\alpha'}{4}h \right)^{-1} \,, \label{eq:RelationHadnX}
\end{align}
\end{subequations}
where we implicitly assumed that the inverse of $Q$ exists, \textit{i.e.} that $h \ne 4/\alpha'$. The equation of motion for the dilaton is
\begin{equation}
    \left[\nabla^2 - \frac{1}{4 \alpha'}\left(\frac{2c}{3} - \beta^{\mu}_{\mu} + {\frac{1}{4}}\beta^2 \right)\right] e^{-\phi} = 0 \,.
\end{equation}
and using the identities \eqref{eq:identitiesRicci} we obtain
\begin{equation}
    \left[\nabla^2 - \frac{1}{48 \alpha'}\left(8 D_c - 8 D + 12 x D - 3 x^2 D \right)\right] e^{-\phi} = 0 \,.
\end{equation}
In $D > 2$ this has the form of a Klein-Gordon equation and we cannot say much about the spectrum of the Laplacian. 

In order to proceed we restrict ourselves to solutions with constant dilaton
\begin{equation} \label{eq:ConstantDilaton}
    \partial_\mu \phi = 0 \,.
\end{equation}
The dilaton equation then produces the algebraic constraint
\begin{equation} \label{eq:algebraicCondition}
    (8 - 12 x + 3 x^2) D = 8 D_c \,,
\end{equation}
which must be solved for $D \in \mathbb{Z}_{>0}$. In particular this is possible only if $h$ is constant. Next, let us study the Einstein equations. It is not difficult to verify that the algebraic condition~\eqref{eq:algebraicCondition} together with $\partial_\mu \phi =0$ imply that the Lagrangian vanishes on-shell. We can therefore ignore the terms we obtain varying the determinant $\sqrt{g}$ as well as those which depend on derivatives of $\phi$. We conclude that the only non-trivial contribution to the Einstein equations is generated by the variation
\begin{equation} \label{eq:variationEinstein}
\delta S \propto \int  d^Dx \sqrt{|g|} e^{-2\phi} \left[\delta g^{\mu\nu} (\beta_{\mu\nu} - {\frac{1}{2}} \beta_{\mu \rho}\beta_{\nu}^{\rho}) + \delta \beta_{\mu\nu} (g^{\mu\nu} - {\frac{1}{2}}\beta^{\mu\nu})\right] \,.
\end{equation}
We observe that
\begin{equation} \label{eq:identityVarBeta}
\begin{split}
\delta \beta_{\mu\nu} g^{\mu\nu} & = - \alpha' g^{\mu\nu} (Q^{-1}\delta Q Q^{-1})_{\mu\nu}{}^{\alpha\beta} R_{\alpha \beta} \\
& = - \alpha' \left(1-\frac{\alpha'}{{4}}h\right)^{-1}g^{\mu\nu} (\delta Q)_{\mu\nu}{}^{\alpha\beta} Q^{-1}_{\alpha\beta}{}^{\rho\sigma}R_{\rho\sigma}  \\
& \doteq \frac{1}{{4}}x^2 g_{\alpha \beta}\delta g^{\alpha\beta} \,,
\end{split}
\end{equation}
where we used the identities~\eqref{eq:identitiesRicci} and we discarded the terms proportional to the variations $\delta R_{\mu\nu}$ and $\delta R_{\mu\nu}{}^{\rho}{}_{\sigma}$ because, when integrated by parts, they give rise to terms that vanish when evaluated on our background; in the last step, we evaluated the only term surviving directly on our background. Exploiting the relation \eqref{eq:identityVarBeta} and the identities \eqref{eq:identitiesRicci}, the non-trivial part of the Einstein equation takes the form of an algebraic constraint
\begin{equation}
g_{\mu\nu} (x-2)(x+4)x = 0\,.
\end{equation}
We therefore obtain the solutions
\begin{equation} \label{eq:xBranches}
    x = 0\,, \, h = 0\,, \, D = D_c  \,,\qquad x = 2\,, \, h = \frac{4}{3 \alpha'} \,, \, D = -2 D_c  \,.
\end{equation}
We observe that the first class of solutions is nothing but the Ricci-flat backgrounds. The second class  can be defined provided that the critical dimension $D_c$ is negative. These backgrounds have positive cosmological constant of order $1/\alpha'$,
\begin{equation}
    \Lambda = \frac{D-2}{2} h = -(D_c+1)\frac{4}{3 \alpha'} > 0 \,.
\end{equation}
Finally, the root $x = -4$ does not admit a solution because Eq.~\eqref{eq:RelationHadnX} cannot be inverted and there exist no $h$.

Another simple ansatz can be obtained assuming instead of~\eqref{eq:ConstantDilaton} that the covariant derivative of the dilaton is covariantly constant 
\begin{equation} \label{eq:CovariantlyConstant}
    \nabla_\mu \phi = \ell_\mu \,, \qquad \nabla_\mu \ell_\nu = 0 \,, \qquad \ell_\mu  \ne 0 \,.
\end{equation}
The condition~\eqref{eq:CovariantlyConstant} is quite restrictive. Applying the commutator of the covariant derivatives to $\ell_\mu$, taking the trace and using that the spacetime is an Einstein manifold we obtain
\begin{equation} \label{eq:conditionHL}
    h \ell_\mu = 0 \,,
\end{equation}
which is compatible only with $h=0$, \textit{i.e.}, with Ricci-flat backgrounds. We have therefore $\beta_{\mu\nu} = 0 $ and the dilaton equation reduce to the condition
\begin{equation} \label{eq:conditionNormEll}
    \ell^2 + \frac{c}{6 \alpha'} = 0 \,.
\end{equation}
The terms that produce non-vanishing contributions to the equations of motion evaluated on the background $\beta_{\mu\nu} = R_{\mu\nu} = 0$ are
\begin{equation}
    \delta S \supset \int d^Dx \sqrt{|g|} \left[-\frac{1}{2}g_{\mu\nu} L \, \delta g^{\mu\nu} + 4 e^{-2\phi} \partial_\mu \phi \partial_\nu \phi \,{\delta g^{\mu\nu} } + \frac{1}{\alpha'}e^{-2\phi} g^{\mu\nu} \delta \beta_{\mu\nu} \right] \,,
\end{equation}
where $L$ is the Lagrangian and we left the variation $\delta\beta_{\mu\nu}$ implicit. The contribution of the last term can not be evaluated with~\eqref{eq:identityVarBeta} because now the dilaton coupling is not constant. It is not difficult to identify that the only contributing terms are those proportional to the variation of the Ricci. Integrating by parts we get
\begin{equation}
    e^{-2\phi} g^{\mu\nu} \delta \beta_{\mu\nu} \doteq e^{-2\phi} \alpha' g^{\mu\nu} \delta R_{\mu\nu} \doteq 4 \,\alpha'e^{-2\phi} (\ell^2 g_{\mu\nu} - \ell_\mu \ell_\nu) \delta g^{\mu\nu} \,,
\end{equation}
where the dot on the equal indicates that we are evaluating the symmetric rank-2 tensor entering in the Einstein equations on our background. It is then not difficult to verify that the Einstein equations boils down to the condition \eqref{eq:conditionNormEll}. For non-vanishing central charge $c \ne 0$, the spacetime has then structure of a local product $M_D \simeq \mathbb{R} \times M_{D-1}$ with $M_{D-1}$ Ricci-flat and signature Lorentzian (Euclidean) signature for space-like (time-like) $\ell_\mu$
\begin{equation}
    ds^2 = \text{sign}(\ell^2) \, dz^2 + h_{ab} dy^a dy^b \,, \qquad R[h]{}_{ab} = 0 \,.
\end{equation}
For $c = 0$, the spacetime admits a null covariantly constant vector and it is Brinkmann. A notable class of Ricci-flat Brinkmann spacetimes are the Ricci-flat pp-waves which further require flat transverse geometry.

\subsection{The fate of the EOMs solution beyond the FRG truncation}

In the previous section we performed a brief scan of the landscape of solutions compatible with the theory \eqref{eq:tseytlin_actionV2}. In particular, we found that the infinite tower of higher-derivative corrections allows us to construct exotic configurations with positive cosmological constants. It is then natural to wonder whether this configuration can be trusted or it is a consequence of the loss of information due to the FRG truncation and the deformation to \eqref{eq:tseytlin_actionV2}. As explained in the previous sections, determining the central charge action \eqref{eq:tseytlin_action} using approximated beta functions, we have a discrepancy among the approximated EOMs we extract from \eqref{eq:tseytlin_actionV2} and the approximated Weyl anomaly cancellation condition \eqref{eq:frg_eoms}. However, among the anomaly cancellation conditions we have \eqref{eq:SecondCondition} which is exact, which we repeat here for the ease of the reader,
\begin{equation} \label{eq:Condition}
    \beta_{\mu\nu} + 2\alpha' \nabla_\mu \nabla_\nu \phi = 0 \,.
\end{equation}
We can then use the this condition to determine whether a background which is a solution of the approximated theory can be a solution of the UV theory built with the FRG at infinite truncation order. More precisely, a background which fails to satisfy \eqref{eq:Condition} should be considered an artifact of the truncation.

We are now going to check whether the configurations derived in the previous section fail to satisfy Eq.~\eqref{eq:Condition}. Imposing that the background is Einstein $R_{\mu\nu} = h g_{\mu\nu}$ and using the identities \eqref{eq:RelationHadnX} the condition \eqref{eq:Condition} becomes
\begin{equation} \label{eq:ConditionV2}
    h \left(1-\frac{\alpha'}{4}h \right)^{-1} g_{\mu\nu} =  -   2 \, \nabla_\mu \nabla_\nu \phi \,.
\end{equation}
Metric compatibility then  implies that $ \nabla_\mu \nabla_\nu \phi$ is covariantly constant. If the RHS is vanishing, \textit{i.e.}, $V_\mu = \nabla_\mu \phi$ is covariantly constant, Eq.~\eqref{eq:ConditionV2} implies that $h = 0$ and the allowed solutions are Ricci-flat with constant dilaton. If  $ \nabla_\mu \nabla_\nu \phi$  is not vanishing, $V_\mu$ is not covariantly constant but is annihilated by the commutator of covariant derivatives. Replacing the Riemann, tracing it and using the fact that the manifold is Einstein we obtain a condition similar to \eqref{eq:conditionHL} of the form $h V_\mu = 0 $ which can only be solved by $h = 0$. We conclude that all solution constructed in the previous section are compatible with condition~\eqref{eq:Condition}, except for the one with positive cosmological constant corresponding to the $x = 2$ branch of Eq.~\eqref{eq:xBranches}.

\section{Conclusions}\label{sec:conclusions}

In this paper we have initiated the application of FRG methods to the string worldsheet. The first long-term aim is to import the technology developed in other approaches as means to explore the strong-curvature regime of string theory. To this end, we have derived FRG equations that describe curvature corrections within the worldsheet formulation of bosonic string perturbation theory (ignoring the tachyon and the B-field for simplicity). These corrections arise from the cancellation of Weyl anomalies. In particular, in Sect.~\ref{sec:ws_frg} we have discussed how the cancellation of Weyl anomalies on the worldsheet connects to $\beta$-functions via (local~\cite{Shore:1986hk}) RG flows, and we have obtained effective spacetime equations of motion within a two-derivative truncation of the worldsheet $\sigma$-model. In Sect.~\ref{sec:ws_frg_beyond2} we have extended the analysis to an infinite-dimensional truncation, which involves all orders in $\alpha'$. This truncation is an analog of $f(R)$ in gravity: although it misses infinitely many tensor contractions, it furnishes an analytically tractable setting where some qualitative all-order effects could be studied. In this sense, we treat this first step as a proof of principle of applicability of the method. With the resulting beta functions and spacetime effective action in hand, we ventured to study black-hole-like and Einstein solutions in Sect.~\ref{sec:solutions}, including the structure of $\alpha'$ corrections and their resummation.

The second long-term aim of this work is to establish some grounds on which the scenario of effective asymptotic safety~\cite{deAlwis:2019aud} could be studied. At present, there is neither substantial evidence for this scenario nor many attempts to find it (see however~\cite{Basile:2021euh,Basile:2021krk}); as such, developing a framework to study strong-curvature regimes of string theory is expected to be useful to connect with a putative (nearly) scale-invariant regime, sitting in between the IR effective field theory and the UV stringy regimes.

An important point to emphasize in this regard is that our investigations are restricted to tree-level, in which case various bootstrap results indicate that the cutoff of gravitational effective field theory marks the onset of higher-spin degrees of freedom (see \emph{e.g.}~\cite{Camanho:2014apa, Cheung:2016wjt, Geiser:2022exp, Eckner:2024pqt, Albert:2024yap, Caron-Huot:2024lbf, Cheung:2025tbr}). Therefore, it might be the case that the scenario envisioned by~\cite{deAlwis:2019aud} be only realizable at strong string coupling. This setting is much more mysterious, and has long been suspected to be connected with the microscopic physics of membranes\footnote{It is worth pointing out that FRG methods have been applied to membranes in a similar context~\cite{Romoli:2021hre}.}~\cite{Townsend:1995gp, Witten:1995ex, Townsend:1995kk, Duff:1996aw}. To make progress, a possible way forward would be to employ FRG methods on the worldsheet to extract quantities that are known to be protected by supersymmetry or constrained by dualities~\cite{Bershadsky:1993ta,Antoniadis:1993ze,Bershadsky:1993cx,Green:1997tv,Gopakumar:1998ii,Gopakumar:1998jq,deWit:1999ir,Hohm:2015doa,Hohm:2019ccp,Hohm:2019jgu,vandeHeisteeg:2022btw,Cribiori:2022nke,vandeHeisteeg:2023ubh,Cribiori:2023sch,vandeHeisteeg:2023dlw,Castellano:2023aum,Calderon-Infante:2025ldq,Castellano:2025ljk,Castellano:2025yur,Aoufia:2025ppe,Castellano:2025rvn,Castellano:2026ojb,Zatti:2026fso,Aoufia:2026ztl}.

At any rate, the existence of an effective AS scenario could potentially be of great computational use, even in the eventuality in which the field-theoretic fixed point were non-unitary, acausal or otherwise problematic~\cite{Basile:2025zjc}. This is because the physical RG flow need not reach the fixed point itself for its existence to constrain low-energy physics. Indeed, the closer the physical trajectory passes by the fixed point, the more low-energy couplings are controlled by it. This could eventually allow the implementation of useful approximation schemes in the computation of low-energy observables in more general backgrounds than the ones that are currently under control.

Beyond the specific technical purview of this paper, the cross-fertilization of techniques borrowed from different approaches and research communities can foster a deeper interaction that goes beyond computational methods~\cite{Buoninfante:2024yth}. An example of this has recently taken place in several interactions between Swampland and Bootstrap communities, which highlights the shared aims and underlying questions, points of agreement on the physics and complementarity of techniques. Further expanding these kinds of interactions between communities seeking insight into quantum gravity can result in even more ideas and progress toward this common goal.

\begin{acknowledgments}
{The research of A.P. and F.F. is supported by a research grant (VIL60819) from VILLUM FONDEN. The Center of Gravity is a Center of Excellence funded by the Danish National Research Foundation under grant No. 184. This project has received funding from the European Union's Horizon 2020 research and innovation programme under the Marie Sklodowska-Curie grant agreement No 101299389 and No 101131233. This work is supported by Simons Foundation International~\cite{sfi} and the Simons Foundation~\cite{sf} through Simons Foundation grant SFI-MPS-BH-00012593-11. I.B. is grateful to Harvard University and the Harvard Swampland Initiative for the hospitality during the final stages of this work. The work of M.Z. is supported by the Alexander von Humboldt Foundation. M.Z. acknowledges the hospitality of the LITP at the University of Chicago and the Simons Center summer workshop, where part of this work was developed. M.Z. is grateful to Miriam Gori for continuous encouragement and support. F.F. is grateful to the Max Planck Institute for Physics in Garching, where the initial stages of this project were carried out, and to Perimeter Institute for Theoretical Physics for their hospitality during the completion of this work. Research at Perimeter Institute is supported in part by the Government of Canada through the Department of Innovation, Science and Economic Development Canada and by the Province of Ontario through the Ministry of Colleges and Universities}
\end{acknowledgments}

\appendix
\section{The two-derivative truncation with non-vanishing B-field}
\label{app:B_field}
We extend the two-derivative truncation of Eq.~\eqref{eq:truncated_action}
to the case of a non-vanishing Kalb-Ramond field. We consider
\begin{equation}
\begin{split}
	\Gamma_k = &  \frac{1}{4\pi \alpha'} \int d^2 \sigma \left[ \sqrt{\gamma} \left(G^{(k)}_{\mu \nu}(X) \, \partial_\alpha X^\mu \partial^\alpha X^\nu + \alpha' \, R^{(2)} \, \phi^{(k)}(X) \right)  \right. \\ & \left. \hspace{2.3cm}+ \epsilon^{\alpha\beta} \, B^{(k)}_{\mu \nu}(X) \, \partial_\alpha X^\mu \partial_\beta X^\nu \right] \,.\label{eq:two_deriv_B}
\end{split}
\end{equation}
As above, we do not specify the ghost sector. Although we work in Euclidean signature, where the $B$-term usually carries an imaginary coefficient, we take it to be real. This ensures that the Hessian operator is self-adjoint and has a well-defined spectral problem; see~\cite{Vassilevich:2003xt} for a more detailed discussion.

The $B$-field enters the Hessian through its field strength $H$, and it is convenient to encode it in a new connection with torsion: in particular, acting on the fluctuation $\xi$, the torsion term reads $1/2 \,\epsilon_\alpha\,^\beta \partial_\beta X_0^\nu H^\mu\,_{\nu \rho}$. Following the same steps as in the $B=0$ case, the Hessian is modified to
\begin{equation}
\begin{split}
    	\left(\Gamma_k^{(2)}\right)_{\mu \nu} = \frac{1}{4 \pi \alpha'} (& -G^{(k)}_{\mu \nu} \, \hat{\Box} - R^{(k)}_{\mu \rho \nu \sigma} \partial X_0^\rho \cdot  \partial X_0^\sigma + \frac{1}{2}\alpha' R^{(2)} \nabla_{\mu}\nabla_{\nu}\phi^{(k)} \\& +\frac{1}{4} \partial X_0^{\rho}\cdot \partial X_0^{\sigma} H^{(k)}_{\mu \rho \lambda}H^{(k)\lambda}\,_{\sigma \nu}+ \frac{1}{2}\epsilon^{\alpha \beta}\partial_{\alpha} X_0^{\rho}\partial_{\beta} X_0^{\sigma}\nabla_\mu H^{(k)}_{\nu \rho \sigma} )\,,
\end{split}        
\end{equation}
where $\hat \Box$ is the d'Alembertian built from the torsionful connection above and $\nabla$ acting on $H$ is the target-space Levi-Civita connection. 

We take the regulator to be the optimized cutoff, and a function of $\hat \Box$, and, upon projection onto the truncation, evaluate the functional trace via the heat kernel expansion. The projected Wetterich equation yields the following system for the target space beta functions
\begin{subequations}
\begin{align}
    \beta^{G}_{\mu \nu}&=\alpha' R^{(k)}_{\mu \nu}+\frac{\alpha'}{4} H^{(k)}_{\mu \rho \sigma}H^{(k)}_{\nu}\,^{\rho \sigma} + \frac{\alpha'}{4}\beta_{G}^{\rho \sigma} \left( R^{(k)}_{\mu \rho \nu \sigma}+\frac{1}{4}H^{(k)}_{\mu \rho \lambda}H^{(k)}_{\nu \sigma}\,^{\lambda} \right) \,, \\
     \beta^{\phi}&=  \frac{D-D_c}{6} - \frac{\alpha'}{2} \, \nabla^2 \phi^{(k)} - \frac{\alpha'}{8} \, \beta_G^{\mu \nu} \nabla_\mu \nabla_\nu \phi^{(k)} \,,\\
      \beta^{B}_{\mu \nu}&=-\frac{\alpha'}{2} \nabla^{\rho}H^{(k)}_{\rho \mu \nu}- \frac{\alpha'}{8}\beta_{G}^{\rho \sigma} \nabla_{\rho}H^{(k)}_{\sigma\mu \nu} \,,
\end{align}
\end{subequations}
where we already set $\eta_{\text{gh}} = \beta^{(G)\mu}_{\mu}$ as required by consistency. 

Setting vanishing field strength $H=0$ recovers Eq.~\eqref{eq:frg_beta_functions}; moreover, at leading order in $\alpha'$ the beta functions for $G$ and $B$ reproduce the standard one-loop result found, \textit{e.g.}, in~\cite{Callan:1985ia}, up to the universal dilaton terms, and up to the sign of the terms quadratic in $H$, which is reversed. This is because of the real coefficient of the $B$-term, upon which $H \to iH$. Instead, this agreement does not hold for the dilaton beta function, which misses the term going schematically as $\alpha' H^2$. The reason for this is power counting: the dilaton couples to the worldsheet curvature with an extra power of $\alpha'$, which implies that the terms involving $\phi$ are captured by a one-loop computation, while the $H^2$ contribution only arises at two loops. Reproducing it within the FRG requires enlarging the truncation.

%\end{framed}

% ---------------------------------------------------------------------------------------------------------------------------------
%		Bibliography
% ---------------------------------------------------------------------------------------------------------------------------------

%\printbibliography

%\bibliographystyle{unsrt}

\printbibliography

%\bibliography{t-dual.bib}

\end{document}